# The Gamow Explorer: A gamma-ray burst observatory to study the high redshift universe and enable multi-messenger astrophysics

N.E. White*[a], F.E. Bauer[b], W. Baumgartner[c], M. Bautz[d], E. Berger[e], S. B. Cenko[f], T.-C. Chang[g], A. Falcone[h], H. Fausey[a], C. Feldman[i], D. Fox[h], O. Fox[j], A. Fruchter[j], C. Fryer[k], G. Ghirlanda[l], K. Gorski[g], K. Grant[d], S. Guiriec[a], M. Hart[g], D. Hartmann[m], J. Hennawi[n], D. A. Kann[o], D. Kaplan[p], J. A. Kennea[h], D. Kocevski[c], C. Kouveliotou[a], C. Lawrence[g], A. J. Levan[q], A. Lidz[r], A. Lien[s], T.B. Littenberg[c], L. Mas-Ribas[g], M. Moss[a], P. O'Brien[i], J. O'Meara[t], D.M. Palmer[k], D. Pasham[d], J. Racusin[f], R. Remillard[d], O.J. Roberts[cu], P. Roming[v], M. Rud[g], R. Salvaterra[x], R. Sambruna[f], M. Seiffert[g], G. Sun[y], N. R. Tanvir[i], R. Terrile[g], N. Thomas[c], A. van der Horst[a], W.T. Verstrand[k], P. Willems[g], C. Wilson-Hodge[c], E.T. Young[u], L. Amati[z], E. Bozzo[aa], O.Ł. Karczewski[ab], C. Hernandez-Monteagudoa[ac], R. Rebolo Lopez[ac], R. Genova-Santos[ac], J.A. Rubino-Martin[ac], J. Granot[ad], P. Bemiamini[ad], R. Gil[ad], E. Burns[ae]

[a]Department of Physics, George Washington University, Corcoran Hall, 725 21st Street NW, Washington DC 20052, USA; [b]Pontificia Universidad Católica de Chile, Vicuna Mackenna 4860, 7820436 Macul, Santiago, Chile; [c]NASA Marshall Space Flight Center, Huntsville, AL 35812, USA; [d]MIT Kavli Institute, MIT, 77 Massachusetts Ave, Cambridge, MA 02139, USA; [e]Harvard-Smithsonian Center for Astrophysics, MS-19, 60 Garden St., Cambridge, MA 02138; [f]Astrophysics Division, Goddard Space Flight Center, Greenbelt, MD 20771, USA; [g]Jet Propulsion Lab, 4800 Oak Grove Dr, Pasadena, CA 91109, US; [h]Department of Astronomy and Astrophysics, The Pennsylvania State University, 525 Davey Lab, University Park, PA 16802, USA; [i]School of Physics and Astronomy, University of Leicester, University Rd, Leicester, LE1 7RH, UK; [j]Space Telescope Science Institute, 3700 San Martin Dr, Baltimore, MD 21218, USA; [k]Los Alamos National Lab, P.O. Box 1663, Los Alamos, NM 87545, USA; [l]INAF-Brera, Osservatorio Astronomico di Brera, Via E Bianchi 46, 23807 Merate (LC), Italy; [m]Department of Physics & Astronomy, Clemson University, 118 Kinard Laboratory, Clemson, SC 29634, USA; [n]Department of Physics, University of California, Santa Barbara, Santa Barbara, CA 93106, USA; [o]Instituto de Astrofísica de Andalucía, Glorieta de la Astronomía, E-18008 Granada, Spain; [p]Department of Physics, University of Wisconsin-Milwaukee, P.O. Box 413, Milwaukee, WI 53201; [q]Physics and Astronomy, Radboud University, Houtlaan 4, 6525 XZ Nijmegen, The Netherlands; [r]Department of Physics & Astronomy, University of Pennsylvania, 209 South 33rd St, Philadelphia, PA 19104, USA;[s]University of Tampa, 401 W. Kennedy Blvd., Tampa, FL 33606; [t]W. M. Keck Observatory, 65-1120 Mamalahoa Hwy., Kamuela, HI 96743; [u]USRA, 7178 Columbia Gateway Dr, Columbia, MD 21046, USA; [v]Southwest Research Institute, 6220 Culebra Road, San Antonio, Texas 78238, USA; [x]INAF IASF-Milano, Via Alfonso Corti 12, I-20133 Milano, Italy; [y]California Institute of Technology, 1200 E California Blvd, Pasadena CA 91125; [z] INAF - Osservatorio di Astrofisica e Scienza dello Spazio, Area della Ricerca del CNR,Via Gobetti 93/3, 40129 Bologna, Italy; [aa]University of Geneva, Department of Astronomy, Chemin d'Ecogia 16, 1290, Switzerland; [ab]Polish Space Agency, Prosta 70, 00-838 Warsaw, Poland; [ac]Instituto de Astrofísica de Canarias, C/ Vía Láctea, s/n E-38205 La Laguna, Tenerife, Spain; [ad]Astrophysics Research Center of the Open University, 1 University Rd, Raanana 43107, Israel; [ae]Louisiana State University, Baton Rouge, Louisiana 70803, USA

*newhite@gwu.edu

**ABSTRACT**

The *Gamow Explorer* will use Gamma Ray Bursts (GRBs) to: 1) probe the high redshift universe ($z > 6$) when the first stars were born, galaxies formed and Hydrogen was reionized; and 2) enable multi-messenger astrophysics by rapidly identifying Electro-Magnetic (IR/Optical/X-ray) counterparts to Gravitational Wave (GW) events. GRBs have been detected out to $z \sim 9$ and their afterglows are a bright beacon lasting a few days that can be used to observe the spectral fingerprints of the host galaxy and intergalactic medium to map the period of reionization and early metal enrichment. *Gamow Explorer* is optimized to quickly identify high-$z$ events to trigger follow-up observations with JWST and large ground-based telescopes. A wide field of view Lobster Eye X-ray Telescope (LEXT) will search for GRBs and locate them with arc-minute precision. When a GRB is detected, the rapidly slewing spacecraft will point the 5 photometric channel Photo-z Infra-Red Telescope (PIRT) to identify high redshift ($z > 6$) long GRBs within 100s and send an alert within 1000s of the GRB trigger. An L2 orbit provides > 95% observing efficiency with pointing optimized for follow up by the *James Webb Space Telescope* (JWST) and ground observatories. The predicted *Gamow Explorer* high-$z$ rate is >10 times that of the *Neil Gehrels Swift Observatory*. The instrument and mission capabilities also enable rapid identification of short GRBs and their afterglows associated with GW events. The *Gamow Explorer* will be proposed to the 2021 NASA MIDEX call and if approved, launched in 2028.

## 1. INTRODUCTION

Gamma Ray Bursts (GRBs) are the most luminous explosions in the universe. They have been observed out to redshift $z \sim 9$ at the highest[1] and have great potential as cosmological probes during the period when the intergalactic medium (IGM) underwent a major state change from Hydrogen being neutral to fully ionized, the Epoch of Reionization (EoR)[2]. For a few days GRBs are a bright beacon that, if observations can be made quickly enough, provide backlights for NIR spectroscopy to directly measure material in the host galaxy and the intervening IGM[3]. Long GRBs result from the core-collapse of massive stars and as such their comoving rate is a tracer of star formation in the early Universe[4-6].

In 1997, the *BeppoSAX* observatory discovered that GRBs have a panchromatic afterglow that lasts hours to weeks[7]. The afterglow results from the relativistic jet ploughing into and shock heating the surrounding circumstellar and interstellar material. The *Neil Gehrels Swift Observatory*[8], hereafter *Swift*, has observed many hundreds of GRBs and their afterglows since launch on November 20, 2004, using a combination of a wide field of view GRB detector, and narrow-field X-ray and UV/optical telescopes on a rapidly slewing spacecraft. By rapidly broadcasting arc-second positions, *Swift* enables space and ground-based observatories to follow up each GRB to determine the redshift of the GRB host galaxy and study the afterglow in detail. This highly successful strategy provided the first detailed measurement of the GRB redshift distribution (Figure 1).

Ground based telescopes are required to determine the redshift, but due to inefficiencies in the availability of these telescopes only a third of *Swift* GRBs have measured redshifts[6]. Over its 15-year lifetime to date *Swift* has detected ~9 GRBs around and above redshift 6, of which only a handful have sufficiently high-quality spectra to make a significant measurement of the host galaxy and IGM. A major obstacle in making progress with *Swift* or other planned missions is identifying quickly the rare high redshift GRBs from the much more numerous lower redshift population. This adds both delays and limits resources because large ground-based observatories are not optimized to respond to every GRB. It is notable that the rate of discovery of high redshift GRBs has fallen off in recent years as less large telescope time has been dedicated to GRB follow up.

*Gamow Explorer* will take the next step in fully utilizing GRBs as cosmological probes. It will address key questions on the timing and profile of the EoR, the early formation of metals and star-formation in the high redshift universe. To do this requires increasing the observed high redshift GRB rate by an order of magnitude, and rapidly providing GRB redshift estimates and locations for immediate follow up with R~3000 IR spectrometers. This will be a complete experiment that combines space and ground-based observations from multiple facilities.

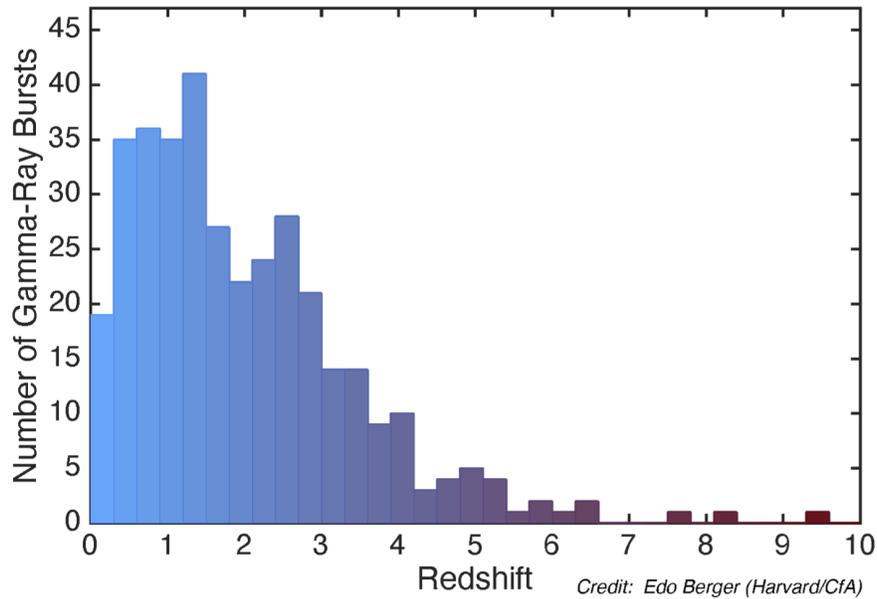

Figure 1: Redshift distribution from 15years of Swift observations combined with redshifts from observatory follow-up.

Short GRBs with durations <2 s is a separate class of GRBs[9] that were proposed to be caused by the merger of binary neutron stars (BNS) or a black hole with a neutron star[10]. This possibility was spectacularly confirmed in 2017 when a Gravitational Wave (GW) event GW170817 was identified concurrently with GRB170817A, which led to the discovery of an associated kilonova (KN)[11-13]. This was the first concurrent detection of an electromagnetic (EM) signal with a GW event. The GRB resulted in the rapid identification of a KN in a relatively nearby galaxy at ~40 Mpc[13]. This spawned a new industry in the study of the formation of the trans-iron elements, the process, and physics of BNS mergers and many associated topics[14].

GRB 170817A was an unusually low luminosity, nearby short GRB and was likely observed slightly off-axis[15]. The current LIGO/Virgo detectors are being upgraded and by 2026 a new, more sensitive generation will come online (the O5/A+ capability) that will push the detection horizon out to 200 Mpc. With only one well studied nearby BNS event there remains much uncertainty in the predicted merger rates and early phase EM emission. *Gamow Explorer* will bring the rapid response capabilities in multi-messenger and time domain astronomy required to locate and study BNS mergers.

The *Gamow Explorer* is named in honor of George Gamow for his seminal contribution to the Big Bang theory of the expanding Universe. The mission will be proposed in late 2021 to the next NASA Medium Class Explorer (MIDEX) opportunity. If approved by NASA the *Gamow Explorer* would launch in 2028. This timing will ensure overlap with JWST during its 10-year design lifetime and the O5/A+ era of ground-based GW detectors.

## 2. SCIENCE GOALS AND OBJECTIVES

The two major science goals for the *Gamow Explorer* are 1) use GRBs to probe the high redshift ($z > 6$) Universe when the first generations of stars were born, galaxies formed, and Hydrogen was reionized; and 2) enable multi-messenger astrophysics by identifying X-ray and Optical-IR counterparts to GW events.

### 2.1 Map the reionization history of the IGM

The EoR is thought to have occurred between redshifts of roughly 6 and 15 (see for example ref. 16), yet the properties of the IGM during this era, and the nature of the sources that drive reionization, remain largely unknown. The most likely scenario is that UV photons emitted by massive stars in the first emerging galaxies are responsible for reionization, but whether small or large galaxies are dominant has been much debated, and Quasi-Stellar Objects (QSOs), X-ray binaries, and dark matter decays are also plausible sources. In a significantly neutral IGM, the damping wing of the Ly-α line imprints a feature red ward of Ly-α at the source redshift which can be used to constrain the neutral fraction of the IGM,

$x_{HI}$, as a function of redshift[17]. This, in turn, provides important input to models of reionization by revealing, for example, whether reionization occurred rapidly or was a more gradual process.

GRB afterglows provide ideal backlights for measuring the IGM damping wing[18]. GRB progenitors reside in typical low mass star forming galaxies, while their high luminosity at early times and featureless power-law spectra makes them perfect backlights, and they are found out to at least $z \sim 9$ when bright QSOs have not yet appeared. GRBs also provide a direct measure of the optical depth of the host galaxy to Lyman continuum radiation and thereby constrain the fraction of UV emission that escapes to reionize the IGM[19].

GRBs provide complementary measurements to other probes of reionization. For example, QSOs may also be used as backlights, however, their spectra are complicated by Lyman-α emission from the vicinity of the QSO, photoionization of the nearby IGM from the central engine and the over-dense environments. QSOs are also found in the most massive galaxies which are not representative of the overall galaxy population. Finally, QSOs are expected to be very rare above $z$ of 8.

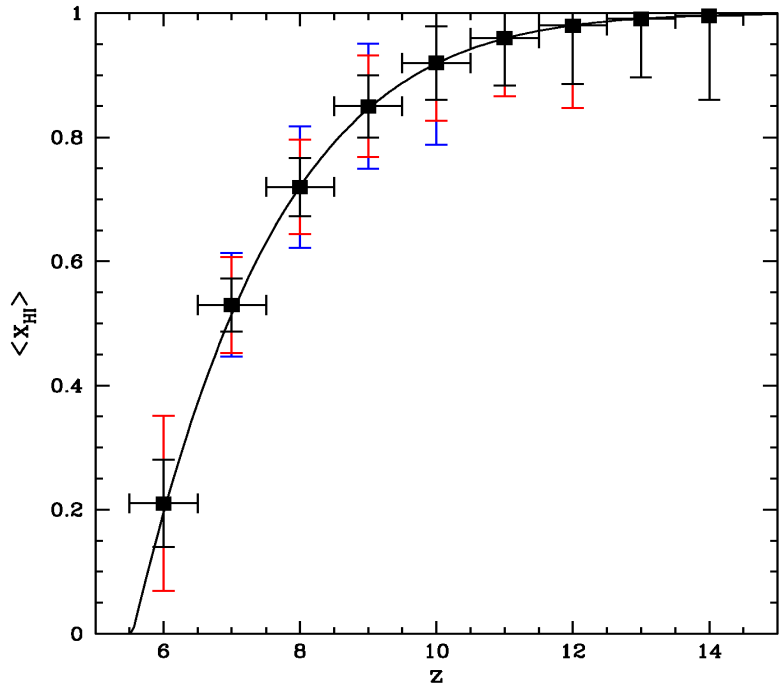

Figure 2. From ref 20. Forecasted constraints on the reionization history of the Universe using a sample of GRBs. The blue points and 1-σ error bars adopt the Gamow Explorer baseline of 20 GRBs, while the red points and error bars show a more optimistic case for 31 GRBs. The black points and error bars show the constraints for an optimistic case of 80 $z \geq 6$ GRBs for the THESEUS mission (which was not selected by ESA for implementation). Each afterglow spectrum is assumed to have an SNR = 20 per R = 3,000 resolution element at the continuum. Note that in the red/blue models the error bars are identical in the $z = 6$ redshift bin since these cases give an identical GRB count in this bin.

Ref. 20 forecasts the reionization history constraints for a sample of $z \geq 6$ GRBs obtained by the *Gamow Explorer*, combined with large aperture ground and space-based spectroscopy. Each model afterglow spectrum is characterized by the size of the ionized region around the GRB host, the volume-averaged neutral fraction outside of the ionized bubble, and the column-density of a local damped Lyman-α absorber (DLA) associated with the host galaxy. A Fisher matrix formalism forecasts the $<x_{HI}(z)>$ constraints obtained from follow-up spectroscopy of GRB afterglows with SNR = 20 per R = 3,000 resolution element. These predictions are shown in Figure 2 for different numbers of GRBs (20, 31, 80). These show that for a baseline sample of 20 GRBs $<x_{HI}(z)>$ can be determined to better than 10-15% (1-σ) accuracy from $z \sim 6$ – 10 in $\Delta z = 1$ bin width. The same measurements can also be used at redshifts 5 to 6 to constrain the escape fraction of ionizing photons from the host galaxy[19]. Taken together, the neutral fraction and escape fraction constraints will help determine the reionization history of the universe and the nature of the sources responsible for reionization[20].

## 2.2 Trace chemical enrichment at high redshift

GRB afterglow spectra often exhibit metal absorption lines, which can be readily discerned redward of the Ly-$\alpha$ damping wing, from heavy elements in the host galaxy and the intervening IGM. The strength of these features (notably ions of C, N, O, S, Mg, Si, Al, and Fe) measured as a function of redshift, probe the chemical enrichment history of the Universe into the EoR and beyond[21-24]. This reflects the cumulative star-formation history from all previous cosmic times, the properties of early stellar populations, the ionization state of enriched gas, and the feedback processes through which galaxies distribute metals into the surrounding gas. Follow-up observations of the host galaxy after the GRB afterglow has faded allow constraints on the mass-metallicity relation during the EoR[25] and provide a powerful test of models of the earliest phases of galaxy formation.

The metallicity levels detectable in GRB afterglows at redshifts $z \sim 6 – 11$ can be estimated by simulating metal absorption spectra of the host galaxies[26]. Each absorption spectrum is parameterized by the hydrogen column density ($N_{HI}$) and metallicity (Z) of the host galaxy, and the neutral gas fraction of the intergalactic medium ($x_{HI}$). The relative abundances of the different metal ions are adopted from those in a mean metal line absorption DLA spectrum[25]. A Monte Carlo sampling is carried out to recover $N_{HI}$ in each of the simulated spectra, as well as the column density of Si II obtained from a curve-of-growth analysis with up to four Si II lines, which provides an estimate of the metallicity and its uncertainty. With a follow-up spectroscopy of GRB afterglows redward of the Ly-$\alpha$, and an expected SNR = 20 per R = 3,000 resolution element, preliminary results suggest that host galaxies with large Hydrogen column density, at log $N_{HI}$ [cm$^{-2}$] >$\sim$ 21.5 yielding strong metal absorption features, can be constrained down to a metallicity of $Z \sim 10^{-3}$ $Z_{solar}$ at $z > 6$ with an uncertainty less than 0.5 dex. On the other hand, low Hydrogen column density systems, at log $N_{HI}$ [cm$^{-2}$] <$\sim$ 19.5 yielding weak metal absorption lines, metallicities down to $Z \sim 10^{-2}$ $Z_{solar}$ can be measured with an uncertainty of less than 0.5 dex.

In addition to measuring the metallicity of host galaxies and its redshift evolution, such measurements, combined with a stellar mass estimate obtained from JWST imaging of the host galaxy after the GRB afterglow fades, allow constraints on the stellar mass-metallicity relation of galaxies during their early phases of formation. Utilizing GRB afterglows as backlights probes the low-mass and low-metallicity limit of galaxies that are likely the typical galaxies responsible for reionization and will illuminate the reionization processes as well as early galaxy formation and evolution.

## 2.3 Survey GRB production at high redshift ($z > 6$)

The comoving density of long GRBs roughly tracks the star formation history of the Universe[5,6]. There are deviations from the SFR history derived from other methods with explanations including: that long GRBs favor lower metallicity environments[27], the fraction of long GRBs relative to the supernova (SN) rate increases with redshift, an evolution in the beaming angle of GRBs with redshift[28] and a change in the underlying IMF of the stellar population[29].

The GRB rate at high redshift provides vital clues to understand the nature of both GRB progenitors and star-formation. There is strong evidence that long-duration GRBs are produced in the collapse of rapidly spinning massive stars, forming an accretion disk around a compact remnant that ultimately drives the GRB jet[29]. In the standard collapsar paradigm[30], the GRB is produced when a disk forms around a black hole and, based on the GRB rate, roughly 10% of such black hole forming stars will result in GRBs. The high-angular momentum needed to produce such a disk and the peculiar nature of the supernovae associated with these bursts (all type Ic) are both difficult to produce in single stars and most progenitor scenarios for long GRBs invoke binary interactions of some kind (e.g., ref. 31). There is evidence that both the fraction of massive stars undergoing binary interactions and the initial mass function of massive stars will evolve at low metallicities and high redshifts. This evolution (and its uncertainties) when incorporated into a range of theoretical models can be used to predict theoretical rates of GRBs at high redshift, that can be compared to observations by the *Gamow Explorer*.

The left panel of Fig. 3 shows theoretical models incorporated in the simulations. For each model, basic calibrations using observed GRBs at low redshift ($z < 4$) from the Swift/BAT data were used. Specifically, the BAT detection numbers for each theoretical model using an approximated BAT detection threshold[32]. The fraction of massive stars with high angular momentum (i.e., the fraction of stars capable of creating GRBs) is adjusted so that each theoretical model matches the number of GRBs produced with the number detected by the BAT[33]. In addition to the total number of GRBs, the BAT GRB redshift distribution estimated from each theoretical model are compared with those from the actual BAT detection[6].

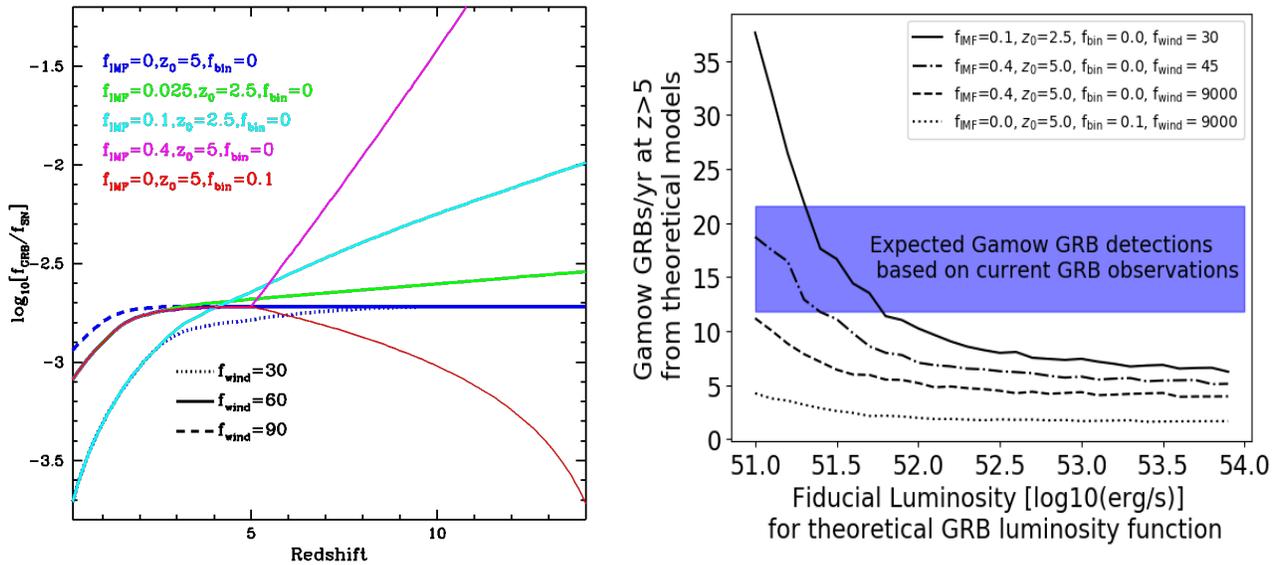

Figure 3 - *Left panel:* A sample of theoretical models with different initial mass functions, binary fractions, and stellar wind parameters. *Right Panel:* Preliminary forecast of the number of *Gamow Explorer* detections per year estimated based on a sample of theoretical models with different luminosity functions (characterized by the fiducial luminosity).

The calibrated samples can be used to estimate the *Gamow Explorer* GRB detections, and study how the *Gamow Explorer* high-$z$ GRB survey will provide insights into stellar evolution at high redshifts. The right panel of Fig. 3 presents a range of *Gamow Explorer* GRB detections based on a sample of theoretical models in this study. With the greatly improved measurements of high-$z$ GRB rate, *Gamow Explorer* will provide crucial information to probe potential changes of the initial mass function of massive stars and star-formation rate at high redshift, and constrain GRB progenitors and their properties (e.g., luminosity distributions, progenitor binary fractions, etc.) in the early Universe.

## 2.4 Search for X-ray and Optical-IR Counterparts to GW Events

The science enabled by GW BNS merger events requires the identification and precise localization of an EM counterpart. Finding such a counterpart quickly is a key goal of *Gamow Explorer*. Such efforts are a major ongoing process worldwide. The future GW error regions are likely to be tens to hundreds of square degrees. As the horizon increases it requires time to search for optical counterparts from the ground both because of the waiting time for the source to become visible from a given observatory, and because of the time required to observe, process the data and to identify and confirm candidates. In contrast *Gamow Explorer* offers a one-stop shop for the identification of such sources. X-rays may be emitted around the time of merger, either in the form of a short-GRB, its afterglow or more isotropic emission from an associated cocoon or central engine (e.g. Magnetar) activity[34]. Recent evidence suggests that BNS mergers create such isotropic emission, visible as either plateau in the X-ray afterglows of short GRBs, or as Fast X-ray Transients (FXRTs). The early identification of EM emission provides unique science (jet physics, evidence of massive NS or BH formation, test of quantum gravity), but also enables rapid observations on the ground to characterize the source, determine redshifts, identify heavy elements, and more.

The rates of such detections remain highly uncertain, not least due to the order of magnitude uncertainties which exist on the cosmic rate of BNS mergers[35]. By 2028 the GW detector network will be operating at a sensitivity that allows it to observe merging neutron stars to well beyond 300 Mpc, corresponding to an observed rate of events a factor of >100 larger than at the time when GW170817 was detected. The advent of additional sensitive detectors via improvements in VIRGO and KAGRA, and the likely addition of a LIGO detector in India will yield smaller error regions[35]. In some cases, advance warnings of a minute or more will provide the ability to be on-source before the merger happens.

# 3. MISSION DESIGN AND IMPLEMENTATION

The *Gamow Explorer* will realize the science goals and objectives by 1) maximizing the number of high redshift GRBs ($z > 6$) detected per year, with at least an order of magnitude more than Swift; 2) rapidly identify $z > 6$ GRBs by having an onboard NIR telescope to identify high redshift candidates; and 3) optimizing the pointing direction to maximize follow up by large ground- and space-based telescopes, including JWST.

The *Gamow Explorer* mission implementation approach builds on the approach pioneered by *Swift*[5]. One instrument, a Lobster Eye X-ray Telescope (LEXT), detects the GRB and provides an arc-minute position. A second, the Photo-z InfraRed Telescope (PIRT), is pointed at this position within 100 s by a rapidly slewing spacecraft to determine an arc second position and whether the GRB is from $z > 6$ within 1000 s of the GRB. A real-time low bandwidth connection to the spacecraft ensures the critical GRB parameters are broadcast to trigger large ground-based telescopes.

## 3.1 Lobster Eye X-ray Telescope (LEXT)

A driving *Gamow Explorer* requirement is to increase the capability to find and identify high redshift GRBs ($z > 6$), by at least an order of magnitude compared to *Swift*. This is achieved by observing in the X-ray band because at high redshifts, the spectral peak of the GRB emission is shifted to lower energies due to the cosmological redshift. It also allows for imaging optics, increasing the signal to noise and position location over coded-mask GRB detectors. The LEXT utilizes an array of slumped micro pore optics (MPOs) with 40 μm square pores[36]. These optics mimic Lobster eyes and provide a very wide field of view imaging in the soft X-ray band[37]. The required performance to detect at least 20 $z > 6$ GRBs over a 3-year prime mission is a combination of field of view (FOV) and focal length, combined with the mass, volume and cost available. A trade study showed that a 30cm focal length combined with a >1000 sq deg field-of-view provides the required performance.

The optics aperture for a LEXT module is formed by a 10 x 5 array of MPOs mounted on a spherical frame with a radius of curvature of 60 cm, giving a focal length of 30 cm. X-rays which reflect off two orthogonal square pore sides, form a central focus (even number of reflections) or a line focus (odd number), giving a cross-arm point-spread-function (PSF), with the remainder flux going through unfocused, creating a diffuse background. In this configuration ~75% of the incident X-rays are focused. There are two LEXT modules with each having an 18.5° x 36.9° FOV, giving a total of ~1350 sq. deg. At the focus of each module is an array of CCDs. Each flat CCD detector is 5 x 5 $cm^2$ and is laid out with 2 x 4 detectors that follow the spherical focal surface and provide an effective pixel size of 45 microns (30 arc-second). The distinctive cross PSF, has a core with a full width half maximum of 7 arc-minute, giving 1-2 arc-minute locations. LEXT covers the 0.3 to 10 keV band. The LEXT instrument team is led by NASA Marshall Space Flight Center (MSFC) with the optics provided by University of Leicester and the focal plane by the MIT Kavli Institute and MIT Lincoln Laboratory. The instrument integration will take place at MSFC, using the stray light facility for testing and calibration. More details can be found in ref. 36.

Predicted long GRB rates for *Gamow Explorer* LEXT were obtained using a population model based on ref 38. This is built through two physically motivated functions describing the GRB luminosity function and redshift distribution. The former is assumed to be a broken power-law extending down to $10^{47}$ erg/s. There is a link between the luminosity and the spectral peak energy[39] so very soft, low luminosity events are also present in the simulated population for the LEXT bandpass. The observed properties of GRB host galaxies[6] suggest a scenario in which the GRB event requires a low metal content of the progenitor, consistent with theoretical expectations[40], with long GRB formation suppressed in high metallicity environments. The GRB formation rate is assumed to be proportional to the cosmic star formation rate multiplied by a factor $(1+z)^\delta$ which increases with redshift because of the decrease in the cosmic metallicity[41]. The free parameters describing the GRB luminosity function, and the value of δ are constrained by reproducing the observer frame properties of bursts detected by the *Fermi*-Gamma Burst Monitor (GBM) and the redshift distribution and rest frame energy/luminosity distributions of bright samples of GRBs detected by *Swift*-Burst Alert Telescope (BAT). To minimize the impact of selection biases, bright flux-limited complete samples are used[38,41]. GRBs are assumed to have a prompt emission spectrum described by the *Band function* whose peak spectral energy is linked to the luminosity or energy through the corresponding *Yonetoku* and *Amati* relationships with the low and high energy spectral slopes sampled from observed distributions. The simulated population events are convolved with the LEXT effective area and the source signal in the 0.3-5 keV energy range is compared with the expected background count rate to select detected events.

The predicted detection rates by the *Gamow Explorer* LEXT instrument, assuming an observational efficiency of 95% and the LEXT instrumental properties are shown in Fig. 4 by the red line with a 1-σ uncertainty (yellow shaded region). For comparison the *Swift* detection rate as obtained with this model (which is calibrated with the *Swift* detected bursts) is shown by the blue line. The predicted $z > 6$ rate for *Gamow Explorer* is between 5 and 10 times the *Swift* detection rate (red and blue line in Fig. 4). The raw *Gamow Explorer* performance relative to *Swift* in detecting GRBs over all redshifts is shown in Fig. 4 by the yellow shaded region.

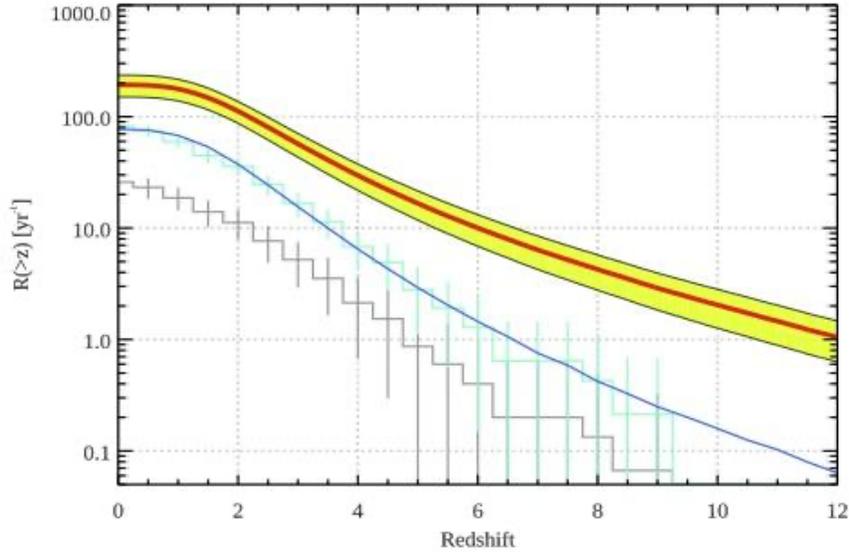

Figure 4: Population model predicted cumulative detection rate of long GRBs by *Gamow Explorer* (red line) and its one sigma uncertainty (yellow region). A model prediction of the *Swift* detection rate is shown by the blue line. The grey histogram shows observed GRBs with measured redshifts and by correcting for an average 30% efficiency of redshift measurement by ground based follow up it is shown to be consistent (cyan histogram) with the *Swift* detection rate curve.

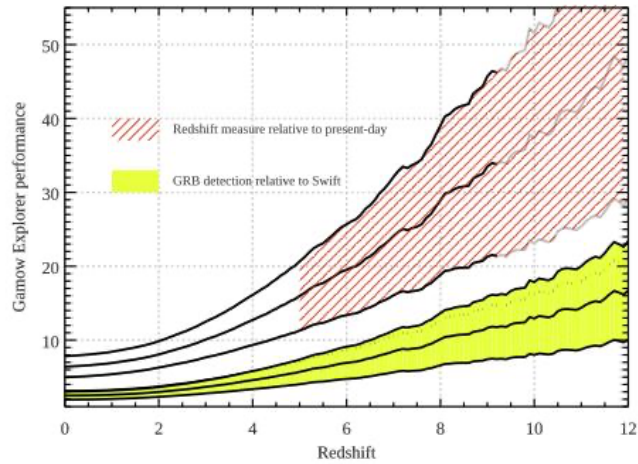

Figure 5: *Gamow Explorer* performance. The yellow filled region shows the cumulative curve, as a function of redshift, of the *Gamow Explorer* performance relative to *Swift* in detecting long GRBs (the shaded region represents the one sigma confidence). The red hatched region shows the *Gamow Explorer* performance in measuring redshifts relative to the present-day performances from ground-based follow up of (mostly) *Swift* detected bursts. Current estimates indicate that ~80% of the high-redshift GRBs detected by the *Gamow Explorer* will have their redshift measured onboard.

For comparison the rate of GRBs with measured redshifts collected in the last 15 years is shown by the grey histogram. By comparing with the expected *Swift* detection rate (blue line), on average, the present ground based follow up redshift recovery of *Swift* detected bursts is ~30% (grey histogram and blue curve). With >80% efficiency of redshift measurement by on-board photometry *Gamow Explorer* will provide about 20 times (between 15 and 25 at 1-σ) more GRBs at $z > 6$ with measured redshifts than the present-day rate. This is shown by the hatched red region shown in Fig. 5. Table 1 shows the *Gamow Explorer* total detection rate for different representative redshift cuts (col.1), the *Gamow Explorer* detection rate with redshifts (col.3) and for comparison of the current redshift measurement rate (col.4). This also illustrates the dramatic improvement that will be obtained with *Gamow Explorer*.

Table 1: Expected annual detection rates of long GRBs by *Gamow Explorer* (with their 1-σ uncertainty interval) corresponding to the red line (and yellow shaded region) in Fig. 4. *The Gamow Explorer* redshift measurement rates correspond to an average redshift recovery efficiency of 80%. The present-day redshift measurement rate is shown for comparison. This is obtained considering the list of bursts detected in the last 15 years which, through ground based follow up with redshift determinations.

| Redshift | Gamow Explorer detection rate [yr$^{-1}$] (one sigma interval) | Gamow Explorer redshift measurement rate [yr$^{-1}$] | Current redshift measurement rate [yr$^{-1}$] |
|---|---|---|---|
| $z > 0$ | 192.6 (149.7 - 235.6) | 154.1 | 25.9 |
| $z > 4$ | 29.6 (21.8 - 37.6) | 23.7 | 2.1 |
| $z > 5$ | 16.6 (11.8 - 21.6) | 13.3 | 0.9 |
| $z > 6$ | 10.0 (6.9 - 13.2) | 8.0 | 0.4 |
| $z > 7$ | 6.2 (4.3 - 8.5) | 5.0 | 0.2 |
| $z > 8$ | 4.2 (2.8 - 5.7) | 3.4 | 0.1 |
| $z > 9$ | 2.8 (1.8 - 3.9) | 2.2 | 0.06 |
| $z > 10$ | 2 (1.3 - 2.8) | 1.6 | 0.0 |

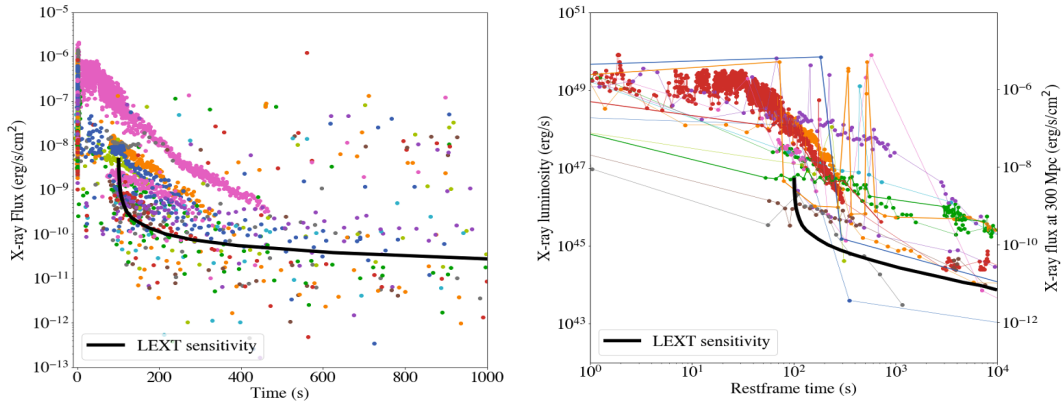

Figure 6: Short GRBs and their afterglows as seen by the Swift BAT and XRT. The Gamow LEXT instrument sensitivity as a function of exposure time is shown as a black line. The left-hand panel shows the short GRB afterglows as observed, while the right hand panel shows them in luminosity space, with their flux at 300 Mpc (a typical distance for second generation detectors at full sensitivity) shown on the right-hand y-axis. The Gamow LEXT is well suited to identify any afterglow-like emission, which it can do for events which are on, or close to the axis of the GRB. Additionally, the long-lived emission seen in some cases has been suggested to be isotropic, so that it may be possible to observe such emission for essentially all mergers.

*Gamow Explorer* LEXT observations will provide X-ray times, positions, fluxes, spectral-slope and absorption measures for the EM counterparts to GW BNS merger events that it identifies. Figure 6 compares Swift afterglow X-ray lightcurves with the LEXT sensitivities, which demonstrates that the afterglow emission from short GRBs is readily detected. This will enable significant science from *Gamow Explorer* alone, as well as when paired with the information which flows from the GW observatories (distances, inclinations, compact object masses). The brightness, lightcurve and spectra of sources as a function of inclination angle provides direct constraints on the jet structure and properties. The times of EM detection relative to the gravitational waves provide a route to testing if the speed of light is equal to the speed of gravity. BNS mergers, the origins of GW events, emit primarily in the NIR[13]. The *Gamow Explorer* PIRT (see 3.2) will obtain early observations of either afterglow or KN; its 5-bands can provide snapshot identification for KN candidates identified in ground-based searches given the unique KN color evolution; finally, it can also track the KN itself, providing continuous visible to IR observations to track heavy element production and the evolution of different KN components.

### 3.2 Photo-z InfraRed Telescope (PIRT)

To create a sample of high redshift long GRBs requires to promptly determine which of the hundreds of GRBs are from high-$z$. The photo-$z$ technique is well proven for rapidly identifying high redshift objects, including GRB afterglows[42, 43]. It takes advantage of the Hydrogen Lyman-α absorption which creates a sharp blue-ward drop out. High redshift GRBs from *Swift* have been identified using this technique from the ground e.g., GRB 090423 at $z \sim 8.2$[3]. To determine the required sensitivity, a sample of high redshift afterglows has been created based on detected afterglows which have been shifted to the required high redshift and the NIR bands (e.g., $z = 6$ and the *J* band), based on the methods presented in ref 44. This is justified as the afterglows of true high redshift GRBs do not differ in luminosity from their lower-$z$ counterparts, and dust extinction appears to be rare at high redshifts. A large sample of known GRB afterglows[45] with additional afterglows which have detections at early times[46,47] has been used. Given the rapid decay typically seen in afterglows at early times, a 500 s exposure is required, beginning within 100 s of the GRB trigger. Figure 7 shows the afterglows of 104 GRBs transformed to $z = 6$ and the *J* band with the time span and flux density level of the first finding chart marked. This demonstrates that a baseline mission limiting sensitivity of 15 µJy (21 mag AB) at 5-σ will allow the detection and determination of the photo-$z$ of at least 80% of the high redshift afterglows within 1000s of the GRB trigger.

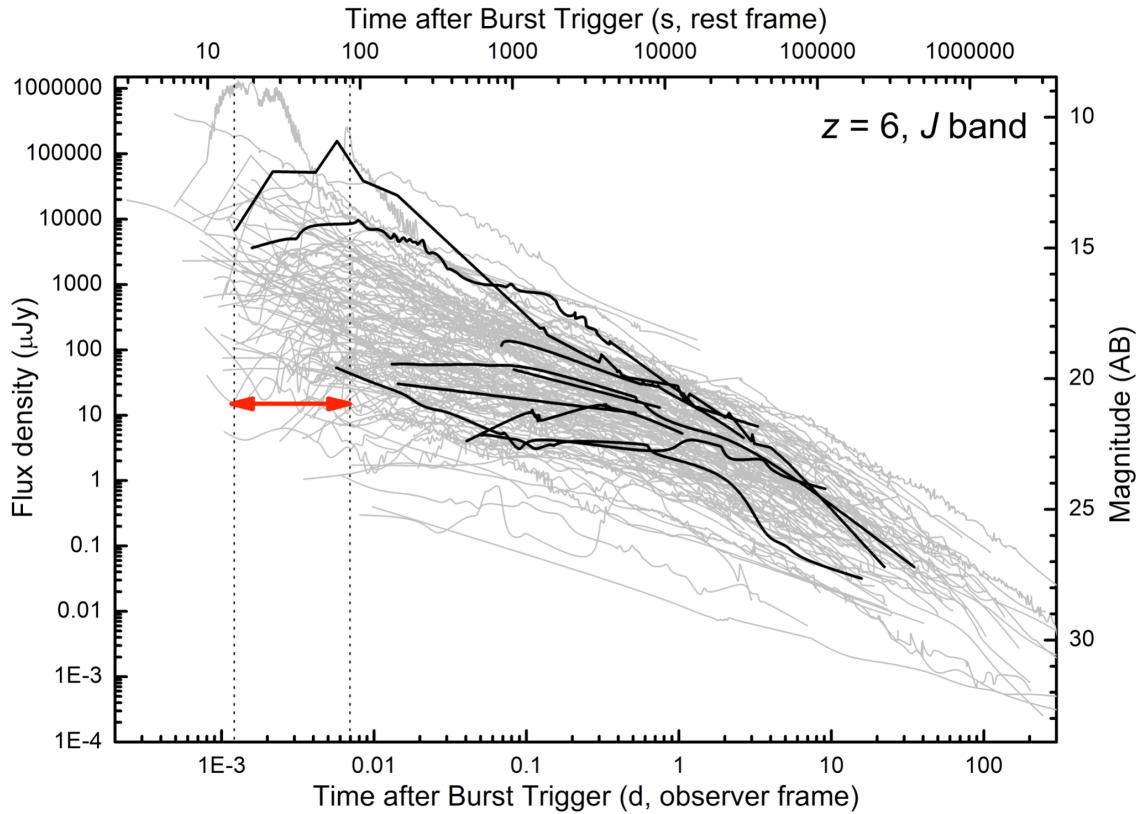

Figure 7. Afterglows of both high-$z$ GRBs (black) and low-$z$ GRBs (grey) shifted to a common redshift of $z = 6$, and into the *J* band. They have been corrected for Galactic (*Gamow Explorer* will observe away from the Galactic plane, and in the NIR) and intrinsic dust extinction (which is very low at high $z$) and are otherwise as observed. The span of the red arrow indicates the 5-σ limiting sensitivity, and the duration of the first PIRT exposure (100 s – 600 s after the trigger), the vertical dotted lines are to guide the eye. Clearly, most early GRB afterglows will be detected at sufficient significance in this first finding chart, allowing a photo-$z$ determination if they lie at high redshift.

A 30 cm aluminum RC telescope design provides the required 15 μJy (21 mag AB) 5-σ sensitivity in 500 s. The field of view is 10 arc min square and is matched to the <2 arc-minute GRB localization radius from the LEXT GRB detections. Five simultaneous channels are obtained by using dichroic prism beam splitters to place five images onto a single detector. The telescope is passively cooled to 200K to avoid thermal emission dominating the sky background in the longest wavelength channel. The focal plane is a flight qualified spare JWST NIRCam H2RG detector, SIDECAR electronics and focal plane assembly. The H2RG detector has 2048 x 2048 pixels, covering the 0.6 to 2.5 micron band and will be passively cooled to 100K. The positional accuracy for the GRB afterglow will be ~1 arc-second. Using this combination, the required sensitivity can be achieved to obtain >80% redshift recovery for GRBs with $z > 6$. The PIRT instrument is led by JPL and more details can be found in ref. 47.

### 3.3 Follow up observations

A key part of the science objectives is to deliver the required NIR medium resolution spectroscopy (R ~ 3000) with signal to noise of ~20 per resolution element, to measure the profile of the Ly-α absorption line and metal absorption lines from the host galaxy[20]. This requires a large collecting area and 6m class or greater telescopes are required. Without *Gamow Explorer* the observatories will not know where and when to point; likewise, without the ground and space-based facilities the science objectives cannot be achieved. This will be a complete end to end experiment that is a partnership between *Gamow Explorer* and the follow up observatories.

The key to obtain the highest quality spectra is to catch the afterglow as early as possible when it is brightest. The *Gamow Explorer* is optimized to send an alert within 1000 s of the GRB trigger to JWST in space and large ground-based facilities including *Keck Observatory*, *Gran Telescopio Canarias* (GTC), *Gemini*, the Very Large Telescope array (VLT) and future planned 30 – 40 m facilities such as E-ELT. The *Keck Observatory* and GTC staff are *Gamow Explorer* team members to help facilitate these rapid follow up observations. By having the capability to identify which of the hundreds of GRBs per year are at $z > 6$, this will remove the current delays and eliminate the need to follow up every GRB with these large facilities. A figure of merit approach will be used to determine which facilities are given priority based on the redshift estimate, the NIR flux, whether it is nighttime at the observatory, weather conditions and observatory operational status. JWST requires 2 days to repoint for a disruptive target of opportunity, but because it is in space with no atmospheric degradation or air glow, it will provide excellent spectra. It will be used for the highest redshift GRBs and those not optimal for observations from the ground because e.g., key features do not fall in the atmospheric wavelength windows.

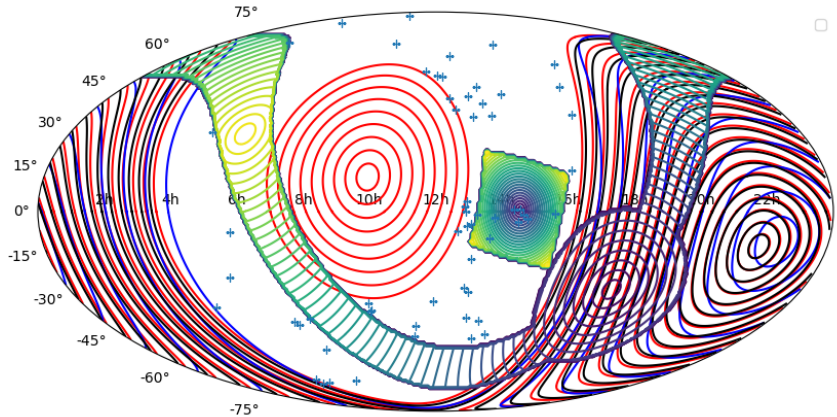

Figure 8. Example visibility for *Gamow Explorer* for the L2 orbit for a simulated day in the life. Contours represent Sun (red), Earth (black), Moon (blue) constraints. Allowed Sun angles of 85° to 135° lead to a constrained region around the anti-Sun, the red contoured region above. Also shown, a region representing a Milky Way avoidance constraint, galaxies from the GLADE catalogue, used as observing inputs, and an example LEXT field of view location (green/yellow/blue square). As *Gamow Explorer* will be at L2, these do not change significantly over an observing day, allowing for long stares.

### 3.4 Orbit and Concept of Operations

We have studied both L2 and low Earth orbits. Based on a JPL Team-X study an L2 halo orbit was selected. An L2 orbit delivers high observing efficiency (> 95%), uninterrupted viewing ideal for finding longer GRBs expected because of cosmological time stretching, immediate follow up by IR telescope for a photo-$z$ measurement, passive cooling of the IR telescope, IR detector and X-ray CCDs, continuous viewing within the JWST field of regard (Sun angle 135° to 85°), optimized pointing for large telescopes in the Canary Islands, Hawaii and Chile, and optimum for avoiding the obscuring galactic plane (which also harbors many bright variable X-ray sources that may generate false alarms). The required GRB alerts arrive via a continuous low-bit-rate (~6000 bps) link from L2 using the KSATLite and other ~4m class ground station dishes, with regular 35m Deep Space Network data dumps for the production data.

The concept of operations is to monitor a patch of the sky waiting for a GRB. When it is detected the LEXT arc-minute position is used to autonomously slew the spacecraft to point the IR telescope at its location within ~100 s. The IR counterpart is identified using a combination of colors (GRBs have a harder spectrum than background galaxies), comparison with catalogues of known sources and time variability. The arc-second location, flux and a photo-$z$ determination are transmitted to the community within 1000 s of the triggers. If the GRB is at $z > 5$ then ground-based telescopes observations are alerted and scheduled in near real time.

### 3.5 Time Domain Astrophysics

During the time when Gamow Explorer is waiting for a GRB it will provide a new capability for Time Domain Astrophysics (TDA). This will be both to utilize the PIRT as a dynamic and flexible resource to respond to and discover transient events in the NIR. The LEXT will serendipitously discover other transient X-ray events, which will have rapid follow up in the NIR by PIRT.

The IR transient sky remains a relatively unexplored frontier but is relevant for many areas of astrophysics. Many sources are bright only in the IR due to opacity or dust or temperature. For example, the *Spitzer* Infra-Red Intensive Transient Survey (SPIRITS) uncovered unusual IR transients with luminosities in between novae and supernovae (SPRITES[31]). Other Spitzer searches revealed that up to 50% of supernovae explode in dusty regions of star forming galaxies and can be obscured by dust at optical wavelengths[39, 40]. Many supernovae form dust at very late times (>1000 days) or heat dust as their shocks interact with surrounding dense circumstellar mediums[41]. Thermonuclear Type Ia SNe are becoming increasingly important standard candles for cosmology in the near-IR (e.g., ref. 42) and late-time detections stand to disentangle the complex radiative transfer process and probe the explosion physics of the progenitor model[43]. While waiting for GRBs the PIRT will conduct a galaxy survey to identify and monitor IR transients based on that pioneered by *Spitzer*. The galaxies will be selected based on their properties and sky locations to be consistent with the prime GRB science requirements. This *Gamow* IR Transient Survey (GIRTS) will be complementary to the optical surveys conducted by the *Rubin* Observatory as well as other ground-based surveys.

The GIRTS provides a pointing and revisiting framework for *Gamow Explorer* that opens powerful applications for TDA science in the soft X-ray band. The LEXT FOV will cover 15-40% of the sky each day, supporting investigations of diverse types of X-ray transients, in parallel with the PIRT survey. The LEXT with its large grasp will serendipitously detect many different classes of X-ray transients including supernova shock breakouts, tidal disruption events, X-ray counterparts to fast blue optical transients, X-ray bright nuclear transients, and flare stars. Of particular interest is the opportunity to partner with the burgeoning industry of optical transients (OT), which numbered 21,600 in 2020[56]. Classifications of OTs are inferred from optical brightness, colors, and variability timescales, but some of the most interesting high-energy objects can avoid recognition until the photometry is supplemented with an X-ray detection. LEXT sky coverage will revolutionize the raster scan and survey efforts ongoing with *Swift* and SRG/e-ROSITA, respectively - selecting the brightest X-ray sources buried among thousands of OTs in the era of the Vera C. Rubin Observatory.

## 4. CONCLUSION

*Gamow Explorer* will be the first observatory dedicated to using GRBs, the most luminous cosmic explosions, to illuminate the era of reionization. It will be a partnership with JWST and the large ground-based telescopes that are required for the required medium resolution NIR spectroscopy to study the Ly-α and metal absorption lines. Gamow will provide reliable alerts as to when a high redshift GRB has occurred. This same rapidly slewing spacecraft/instrument complement will also rapidly locate EM counterparts to GW-emitting mergers and be a key asset for time domain astrophysics observations. The 2028 to 2031 prime mission phase is timely by overlapping the ten-year design lifetime of JWST, the coming era of 30m+ observatories and the advanced GW facilities.